\documentclass[floatfix,rmp,twocolumn,twoside]{revtex4}
\usepackage{graphicx}

\bibpunct{[}{]}{,}{n}{}{,}

\usepackage[english]{babel}
\usepackage{verbatim}
\usepackage{subfigure}
\usepackage{varwidth}
\usepackage{graphicx}
\usepackage{alltt}

\makeatletter
\g@addto@macro\@verbatim\small
\makeatother

\begin{document}

\title{Treating Insomnia, Amnesia, and Acalculia in Regular Expression Matching}
\author{Luis~Quesada, Fernando~Berzal, and Francisco J.~Cortijo\\
  Department of Computer Science and Artificial Intelligence, CITIC, University of Granada, \\
  Granada 18071, Spain \\
  \textit{lquesada@decsai.ugr.es, fberzal@decsai.ugr.es, cb@decsai.ugr.es}
  }

\begin{abstract}

Regular expressions provide a flexible means for matching strings and they are often used in data-intensive applications.
They are formally equivalent to either deterministic finite automata (DFAs) or nondeterministic finite automata (NFAs). Both DFAs and NFAs are affected by two problems known as amnesia and acalculia, and DFAs are also affected by a problem known as insomnia.
Existing techniques require an automata conversion and compaction step that prevents the use of existing automaton databases and hinders the maintenance of the resulting compact automata.
In this paper, we propose Parallel Finite State Machines (PFSMs), which are able to run any DFA- or NFA-like state machines without a previous conversion or compaction step. PFSMs report, online, all the matches found within an input string and they solve the three aforementioned problems.
Parallel Finite State Machines require quadratic time and linear memory and they are distributable. Parallel Finite State Machines make very fast distributed regular expression matching in data-intensive applications feasible.

\end{abstract}

\maketitle

\section{Introduction}

A regular expression, commonly called regex, provides a flexible means for matching strings \cite{Aho1991} often used in data-intensive applications such as lexical analysis in language processors \cite{Levine1992,Quesada2011a}, XML tokenization \cite{Hosoya2001}, bibliographic search \cite{Aho1975}, database queries \cite{Date1990}, spam filters \cite{Xie2008}, virus detection \cite{Lee2007}, network intrusion detection \cite{Snort}, and other data mining applications \cite{HanJ2006,Gomez2008}.

The implementation of regex matching relies on either deterministic finite automata (DFAs) or nondeterministic finite automata (NFAs) \cite{Hopcroft2007}.

Both DFAs and NFAs suffer from amnesia and acalculia, and DFAs also suffer from insomnia \cite{Kumar2007}. Amnesia is the inability to consider the progress of multiple partial matches. Acalculia is the inability to find subexpression occurrences. Insomnia is the inability to unload from memory, or set to sleep, rarely used big portions of automata when they are not needed. 

Several techniques exist that solve those problems by converting sets of regular expressions into compact state machines \cite{Pasetto2010,Becchi2007,Kumar2006,Ficara2008}.
The required conversion or compaction step prevents using existing automata that could be already available in antivirus signature databases or complex data filters. It also makes the modification of the set of regular expressions more difficult, as the conversion and compaction steps have to be performed every time they are modified.

In this paper, we propose Parallel Finite State Machines (PFSM).

A Parallel Finite State Machine is an automaton that can have multiple active states. It can be used for efficiently finding all the matches of a set of regular expressions in a given input string, and it solves the amnesia and acalculia problems.
They also mitigate the effect of insomnia by reducing the number of states in the resulting automaton.

Parallel Finite State Machines do not require conversion and compaction steps and they can, therefore, be run on existing DFAs or NFAs.
They also allow the efficient addition or removal of regular expressions from existing automata.

Parallel Finite State Machines can perform regular expression matching in quadratic order of efficiency and have linear memory space requirements, apart from automata storage, in the worst case.
Moreover, they allow the parallelization of the regular expression matching process with almost linear scalability in the practice.

\section{Background}

A DFA is a tuple $(\Sigma, Q, q_0, \delta, F)$, where:
\begin{itemize}
\item $\Sigma$ is the input alphabet (i.e. a finite, non-empty set of symbols).
\item $Q$ is a finite, non-empty set of states.
\item $q_0$ is an initial state, an element of $Q$.
\item $\delta$ is the state-transition function: $\delta: Q \times \Sigma \rightarrow Q$.
\item $F$ is the set of final states, a subset of $Q$.
\end{itemize}

In a NFA, the state-transition function would be $\delta: Q \times \Sigma \rightarrow \mathcal{P}(Q)$. In other words, $\delta$ returns a set of states.

NFA-like state machines can be converted into DFA-like machines by expanding their states and removing any nondeterminism \cite{Sipser2005}. Each state in the DFA will correspond to a set of states in the NFA. Although DFAs are faster than NFAs, there can be up to $O(2^n)$ states in a DFA, being $n$ the number of states in the NFA.

Amnesia can be inefficiently solved by using a separate state for each combination of every potentially simultaneous partial match \cite{Kumar2007}. Somehow, this is analogous the approach of converting NFAs into DFAs, but much more memory-intensive and therefore, quite impractical.

Acalculia can be solved by repeating the whole regex matching process for each input string index as a starting position and stopping at each possible ending position \cite{Quesada2011a}. This solution is time-intensive and does not allow for the progress of partial matches to be simultaneously considered, thus preventing the solution of amnesia.

When the number of states in an automaton increases, as a result of converting a NFA into a DFA or considering simultaneous partial matches, memory use skyrockets and insomnia has to be taken into consideration.
Insomnia can be solved by swapping rarely used portions of automata out to the hard drive when they have not been used for a while \cite{Kumar2007}. However, that approach presents two major drawbacks. When those portions are used back again, there might be a swapping delay. Also, automata might be so huge that even swapping it to hard drive could be impractical. The most common approach in practice consists of avoiding the expansion of states \cite{Kumar2006,Ficara2008,Becchi2007}, thus eliminating the major cause of insomnia.

DotStar \cite{Pasetto2010} solves amnesia and acalculia by converting sets of regular expressions into compact state machines pertaining to a subclass of DFA with status bits associated to their states.

The technique proposed in \cite{Becchi2007} reduces the memory requirement of DFAs by merging non-equivalent states and labelling their input and output transitions.

D${}^2$FAs \cite{Kumar2006} reduce the memory requirement of DFAs by assuming default transitions for all the states. Default transitions differ from epsilon transitions in that, in case a state cannot find a transition for a specific input symbol, the default transition will be followed and a transition with that specific input symbol will be looked for in the target state. Several default transitions could be followed for processing a single symbol.

$\delta$FA \cite{Ficara2008} is an extension of D${}^2$FA that determines some of the results of chaining default transitions and non-default transitions and makes them explicit in the automaton for optimization. It also uses local memory to store the set of transitions going out from a state that is the source of a default transition. If the default transition of the target state brings to the source state, the transition set is known without having to follow the default transition back to the source state.

These techniques require a conversion or compaction step that makes them unable to run on existing uncompacted DFA- or NFA-like state machines that could be already available in automaton databases.
Moreover, compaction forbids regular expressions to be added or removed from the state machine directly. Therefore, compaction hampers the maintainability of the automata.
It should be noted that the only way to add or remove regular expressions from the state machine would involve converting and compacting the whole new regular expression set, which is costly in terms of time when the set of regular expressions is complex.

Lamb \cite{Quesada2011a} is a lexical analyzer that partially solves acalculia by greedily matching every pattern starting at every position of the input string.
However, it does not find all the possible submatches, as greedily-matched regular expressions find the longest possible matching and discard any shorter matchings.
That is, even though Lamb finds matchings starting at any position in the input string, it does not finds matchings ending at different positions in the input string for the same starting position. This implies that some matchings may indeed be missed (i.e. Lamb still suffers from acalculia).

\section{Parallel Finite State Machine}

In order to solve the amnesia, acalculia, and insomnia problems, we propose the use of Parallel Finite State Machines (PFSMs).

\subsection{Definition}

A PFSM is a concurrent automaton in which several states may be active at the same time and final states include associated labels.
This feature enables matching an input string at several starting positions with a set of regular expressions in order to find all the possible matches.
A PFSM generalizes all existing combinations of DFA- and NFA-like state machines.

A PFSM is a tuple $(\Sigma, Q, q_0, \delta, F, L, l)$, where:
\begin{itemize}
\item $\Sigma$ is the input alphabet.
\item $Q$ is a finite, non-empty set of states.
\item $q_0$ is an initial state, an element of $Q$.
\item $\delta$ is a NFA-like state-transition function: $\delta: Q \times \Sigma \rightarrow \mathcal{P}(Q)$.
\item $F$ is the set of final states, a subset of $Q$.
\item $L$ is the set of labels that identify the different regular expressions.
\item $l$ is the state-label function: $l: F \rightarrow L$.
\end{itemize}

A set of DFAs representing different regular expressions can be put together by using an ancillary initial state with epsilon transitions to each of the initial states of the different DFAs, and by assigning each final state a label identifying that identifies the regular expression it represents.
A PFSM also allows trading off processing time for memory use by considering NFAs instead of DFAs for some regular expressions.

\begin{figure}[htb*]
\centering
\includegraphics[scale=0.88]{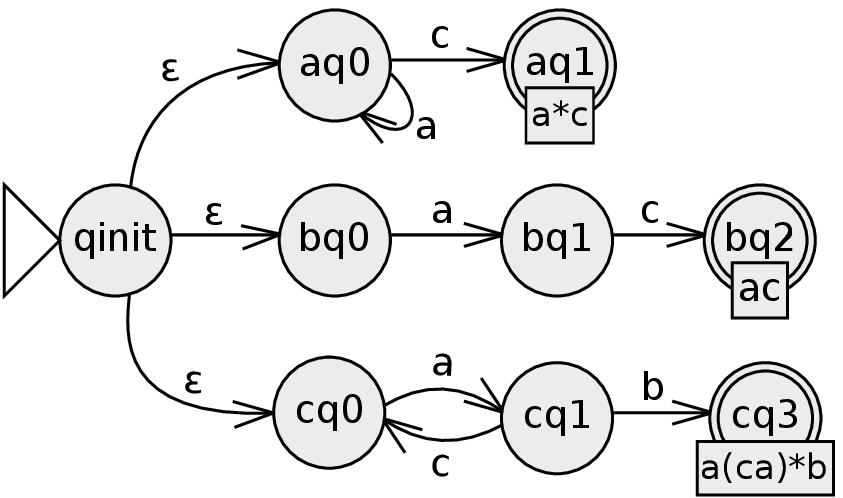}
\caption{A PFSM that compiles the ``a*c'', ``ac'', and ``a(ca)*b'' regular expressions.}
\label{fig:fcsmexample}
\end{figure}

For example, an automaton that considers the regular expressions ``a*c'', ``ac'', and ``a(ca)*b'' is shown in Figure \ref{fig:fcsmexample}.
When trying to match the input string ``aacacab'', that automaton returns all the matches shown in Figure \ref{fig:matches}.

\begin{figure}[htb!*]
\centering
\begin{varwidth}{\linewidth}
\begin{verbatim}
                  INPUT: "aacacab"
    "a*c" matches [0-2]: "aac"
    "a*c" matches [1-2]:  "ac"
     "ac" matches [1-2]:  "ac"
    "a*c" matches [2-2]:   "c"
    "a*c" matches [3-4]:    "ac"
     "ac" matches [3-4]:    "ac"
    "a*c" matches [4-4]:     "c"
"a(ca)*b" matches [1-6]:  "acacab"
"a(ca)*b" matches [3-6]:    "acab"
"a(ca)*b" matches [5-6]:      "ab"
\end{verbatim}
\end{varwidth}
\caption{Results of matching the ``aacacab'' input string using the PFSM in Figure \ref{fig:fcsmexample}.}
\label{fig:matches}
\end{figure}

As the whole automaton does not need to be converted into a single DFA, it does not have to be expanded with too many states, which cuts down its memory requirements, thus mitigating the impact of insomnia.

Also, as amnesia is solved by allowing several simultaneously active states that can represent parallel partial matchings, it is unnecessary to expand the automaton with states representing different simultaneous partial matchings. This drastically reduces the impact of insomnia.

Furthermore, as the automaton does not need to be compacted, regular expressions can be efficiently added or removed from a PFSM. This makes PFSM maintenance easier.

\subsection{Algorithm}

A PFSM is not only initialized once, by activating the initial state at the start of the input string. It is also reinitialized every time an input symbol is going to be processed, in order to consider simultaneous partial matchings starting at different positions in the input string.

Whenever the automaton is initialized or reinitialized, the initial state is activated and it is annotated with a tag that represents the current input string position, which will be used when a match is completed in order to know the starting position of the matched substring. The tag will be propagated unmodified to other states when applying transitions, and a state can include several tags at any given time, which will correspond to multiple simultaneous matches being in the same state while starting at different input string positions.
It should be noted that, whenever the automaton is reinitialized, any active states are kept active.

Using a PFSM, each input symbol is processed in four steps:
\begin{enumerate}
\item The automaton is initialized or reinitialized by activating the initial state and tagging it with the current input string position.
\item All epsilon transitions from the active states (e.g. the initial ancillary state) are iteratively applied and the target states of those transitions are tagged.
\item The input symbol is consumed and, from all the transitions from the active states, only those that match the current symbol are traversed.
\item The traversed transitions are applied and the target states of those transitions are tagged.
\end{enumerate}

\begin{figure*}[p]
\centering
\subfigure[Initialization (input{[0]} is 'a').]{
\includegraphics[scale=0.88]{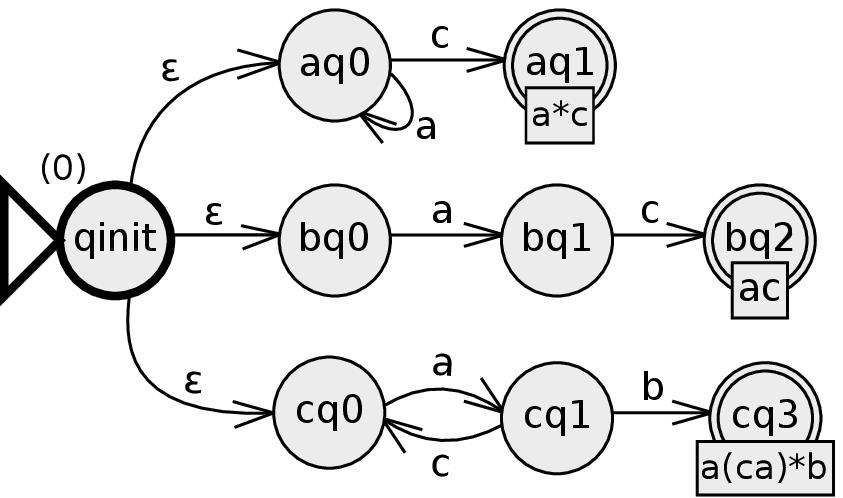}
\label{fig:runninga}
}
\subfigure[Application of epsilon transitions.]{
\includegraphics[scale=0.88]{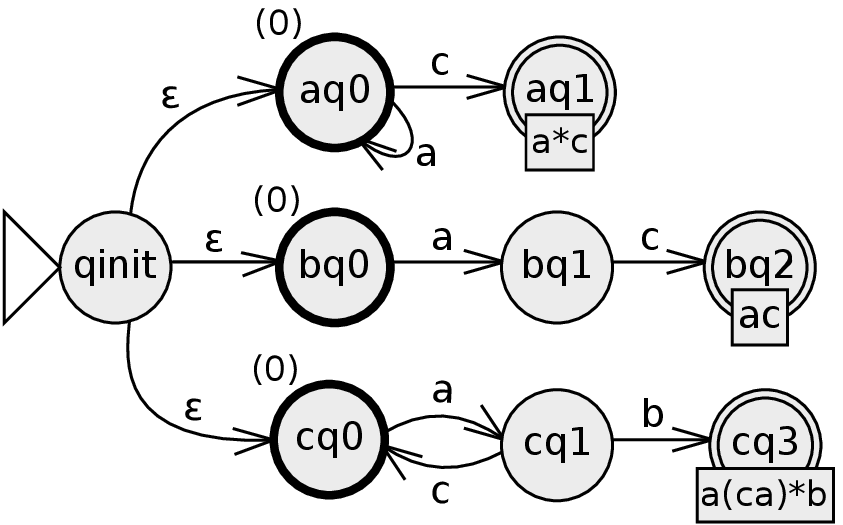}
\label{fig:runningb}
}

\subfigure[Transitions matching 'a'.]{
\includegraphics[scale=0.88]{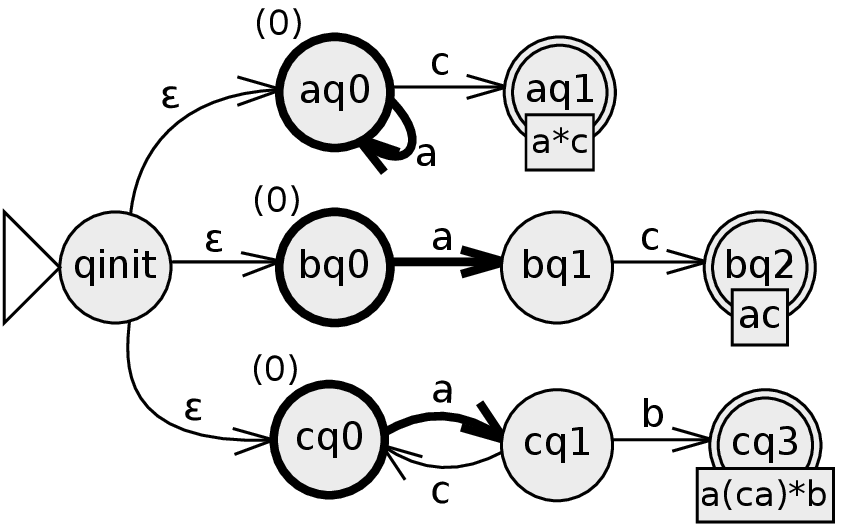}
\label{fig:runningc}
}
\subfigure[Application of transitions.]{
\includegraphics[scale=0.88]{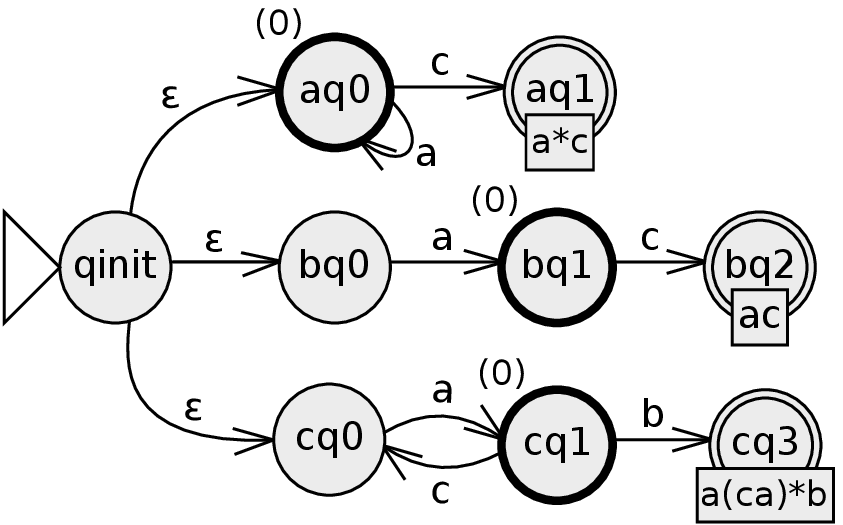}
\label{fig:runningd}
}

\subfigure[Reinitialization (input{[1]} is 'c').]{
\includegraphics[scale=0.88]{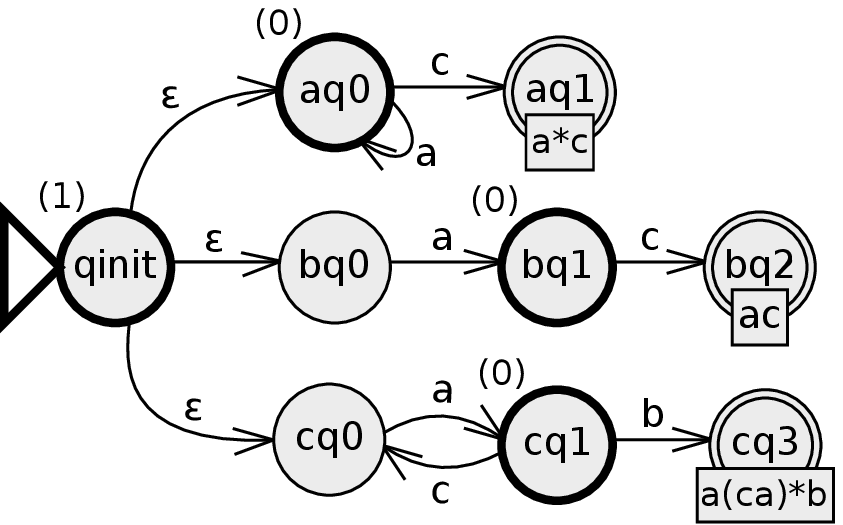}
\label{fig:runninge}
}
\subfigure[Application of epsilon transitions.]{
\includegraphics[scale=0.88]{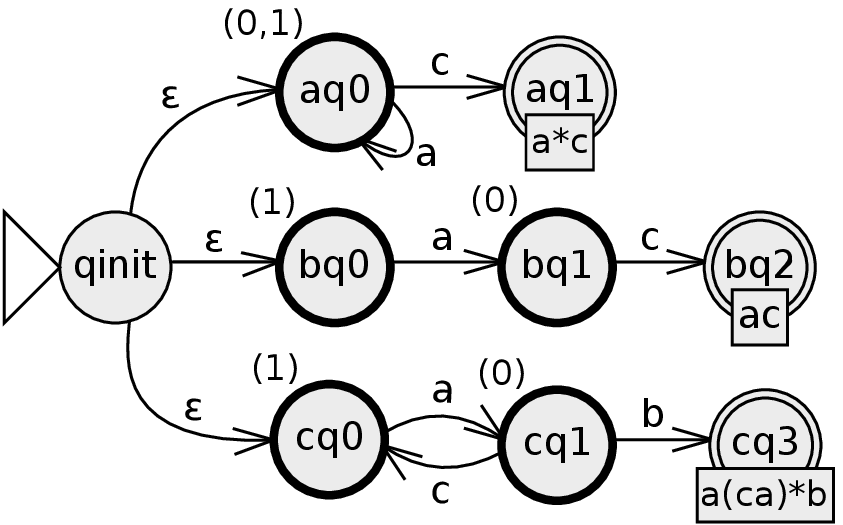}
\label{fig:runningf}
}
\subfigure[Transitions matching 'c'.]{
\includegraphics[scale=0.88]{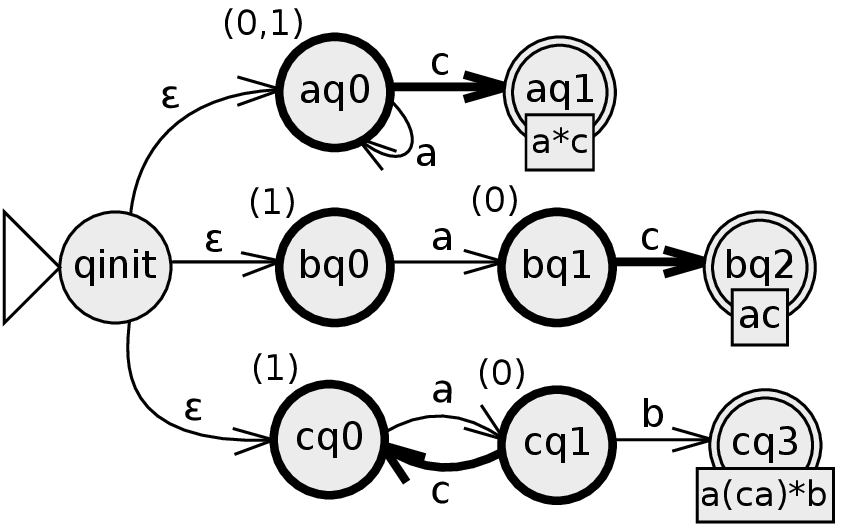}
\label{fig:runningg}
}
\subfigure[Application of transitions. Three matches ending at index 1 are found.]{
\includegraphics[scale=0.88]{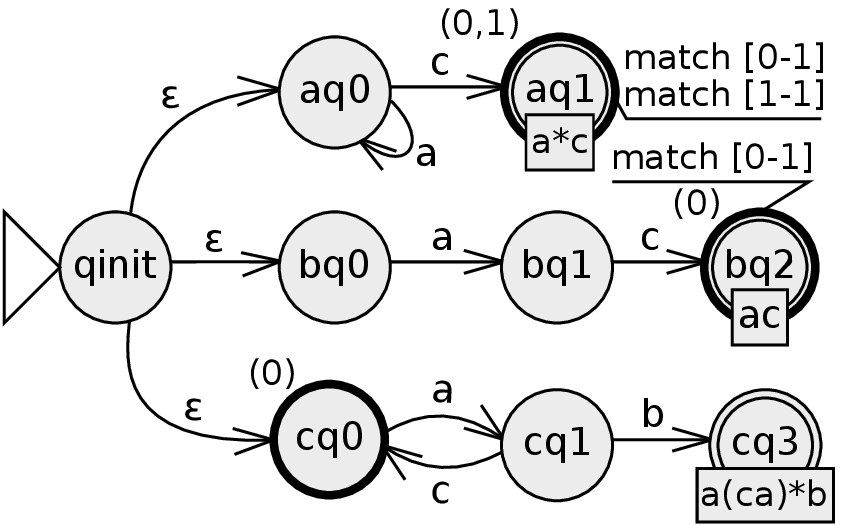}
\label{fig:runningh}
}
\caption{Example PFSM running for two full cycles. The final state labels specify which regular expression that state corresponds to. Active states and transitions are shown in bold. The numbers between parentheses are the valid starting positions for the states.}
\label{fig:runningfcsm}
\end{figure*}

Figure \ref{fig:runningfcsm} shows how the PFSM from Figure \ref{fig:fcsmexample} consumes the two symbols in the ``ac'' input string.

Whenever a transition is applied, in the second or fourth step of the algorithm above, the target state is checked. When it is final, a matching has been found. The matching starting position is given by the final state tag (or tags). The matching ending position is the current input string position. The matched regular expression is identified by the label of the final state.

\subsection{Efficiency analysis}

Given an alphabet with $s$ symbols and $r$ regular expressions, being $m$ the number of states of an automaton representing a regular expression, a PFSM will contain $mr+1$ states.
A maximum of $n^2r$ matchings will be performed in the input string.

If regular expressions are implemented as DFAs, the resulting PFSM automaton may contain up to $mr$ states and $mrs$ transitions.
At most, $r$ states may be active when considering a single symbol of the input string.
As matchings starting at different input string positions are considered, $O(nr)$ states may be active in the PFSM at any given time.
The PFSM processes one symbol at a time, so the process is repeated for the $n$ symbols in the input string.
Therefore, PFSMs' efficiency is $O(n^2r)$ in time and $O(nr+mrs)$ in space.

If regular expressions are implemented as NFAs, the resulting PFSM automaton may contain up to $mr$ states and $m^2rs$ transitions.
At most, $mr$ states may be active when considering a single symbol of the input string.
As matchings starting at different input string positions are considered, $O(nmr)$ states may be active in the PFSM at any given time.
The PFSM processes one symbol at a time, so the process is repeated for the $n$ symbols in the input string.
Therefore, PFSMs' efficiency is $O(n^2mr)$ in time and $O(nmr+m^2rs)$ in space.

Our suggested implementation of PFSMs allows trading off time for space by expressing some of the regular expressions as NFAs instead of DFAs.

It should be noted that PFSMs, as described here, can be implemented as Moore-like machines. A Mealy-like implementation would yield no reduction on the number of states and would decrease the efficiency of the PFSM implementation both in terms of time and space.

\subsection{Parallelization}

The PFSM is a parallelizable automaton with an almost linear scalability.

\subsubsection{Regular Expression Partitioning}

PFSMs can be partitioned by distributing the set of regular expressions among the different processors.
Each processor will find the matches within a subset of regular expressions, which reduces the automaton size at a given processor.
By following this approach, the memory use at a given processor can be reduced just to the space that is necessary to store a single regular expression.
Although the idea of PFSM partitioning is trivial, the fact that PFSM does not compact the regular expression set that allows this kind of parallelization on the fly, in contrast to existing techniques that need to distribute the regular expression set and compact each subset separately.

Furthermore, when some regular expressions are expressed as NFAs in order to save memory, it is possible to distribute both NFA and DFA regular expressions among processors in such a way that memory use and processing time is balanced in each processor, as proposed in \cite{Sun2010}.

\subsubsection{Data Partitioning}
PFSMs can be partitioned by distributing the input string among the different processor. Two different approaches are proposed:

\begin{itemize}
\item Lazy data partitioning. The input string is distributed among the different processors. Each processor will find the matches that start within its segment of the input string and end either in or after that segment.
To achieve this, the automaton reinitialization is disabled after surpassing the ending position of the segment, but the processing does not stop until the end of the input string is reached or there are no active states.
Each processor may need to ask other processors for their segment in order to finish its processing.

\item Chained data partitioning. The input string is distributed among the different processors. Each processor will find the matches that start and end within its segment. The tagged active states at the end of the processing are sent to the next processor, a different set of tagged active states are received from the previous processor and the whole cycle is repeated (without reinitializations) until no more tagged active states are received from the previous processor.
\end{itemize}

Data partitioning linearly reduces memory use with respect to the number of processors.
Assuming a maximum token length of twice the segment length, data partitioning also reduces processing time linearly with respect to the number of processors.

\section{Conclusions and Future Work}

Regular expressions provide a flexible means for matching strings and they are often used in data-intensive applications.

The implementation of regular expression matching typically relies on deterministic finite automata (DFAs) or nondeterministic finite automata (NFAs).
Both DFAs and NFAs are affected by amnesia and acalculia, and DFAs also suffer from insomnia.

Techniques exist that solve those problems by converting sets of regular expressions into compact state machines. This approach, however, presents two major drawbacks: it prevents the use of existing automata that could be already available in antivirus signature databases or complex data filters; and it hinders the maintenance of the resulting automata, since the whole set has to be converted and compacted again whenever a regular expression has to be added or removed from the set.

We have proposed Parallel Finite State Machines (PFSMs), which allow multiple active states and efficiently find all the matches of a set of regular expressions in an input string, solving amnesia and acalculia. PFSMs also mitigate the effect of insomnia by reducing the number of states in the resulting automaton.

As PFSMs do not require any conversion or compaction step, they are able to run on existing DFA- or NFA-like machines, and they allow the addition or removal of regular expressions with zero downtime, making the maintenance of the automaton easier.

PFSMs can perform regular expression matching in quadratic time and have linear memory space requirements, apart from automata storage, in the worst case.
On top of that, PFSMs support three different approaches for parallelization with almost linear scalability in the practice.

Therefore, PFSMs make very fast distributed regular expression matching in data-intensive applications feasible.

We plan to apply PFSMs to cure Lamb \cite{Quesada2011a} from its partial acalculia.
We also plan to apply PFSMs to other data-intensive applications.



\bibliographystyle{plain}
\bibliography{doc}

\end{document}